\documentclass[12pt,manuscript,doublespace]{aastex}

\usepackage{natbib}                
\usepackage{verbatim}
\usepackage{subfigure}
\usepackage{epsfig}

\def\arcsec{\ifmmode '' \else $''$\fi}

\def\arcsecpoint{\ifmmode ''\!. \else $''\!.$\fi}

\def\kms{\ifmmode {\rm km\ s}^{-1} \else km s$^{-1}$\fi}
\def\Msun{\ifmmode {\rm M}_{\odot} \else M$_{\odot}$\fi}
\def\Lsun{\ifmmode {\rm L}_{\odot} \else L$_{\odot}$\fi}
\def\Zsun{\ifmmode {\rm Z}_{\odot} \else Z$_{\odot}$\fi}

\def\ergscm2{ergs\,s$^{-1}$\,cm$^{-2}$}
\def\icm3{{\rm cm}^{-3}}
\def\icm2{{\rm cm}^{-2}}
\def\qo{\ifmmode q_{\rm o} \else $q_{\rm o}$\fi}
\def\Ho{\ifmmode H_{\rm o} \else $H_{\rm o}$\fi}
\def\ho{\ifmmode h_{\rm o} \else $h_{\rm o}$\fi}

\def\vFWHM{\ifmmode v_{\mbox{\tiny FWHM}} \else
            $v_{\mbox{\tiny FWHM}}$\fi}
\def\CCF{\ifmmode F_{\it CCF} \else $F_{\it CCF}$\fi}
\def\ACF{\ifmmode F_{\it ACF} \else $F_{\it ACF}$\fi}
\def\Halpha{\ifmmode {\rm H}\alpha \else H$\alpha$\fi}
\def\Hbeta{\ifmmode {\rm H}\beta \else H$\beta$\fi}
\def\Hgamma{\ifmmode {\rm H}\gamma \else H$\gamma$\fi}
\def\Hdelta{\ifmmode {\rm H}\delta \else H$\delta$\fi}
\def\Lya{\ifmmode {\rm Ly}\alpha \else Ly$\alpha$\fi}
\def\Lyb{\ifmmode {\rm Ly}\beta \else Ly$\beta$\fi}
\def\Lyg{\ifmmode {\rm Ly}\beta \else Ly$\gamma$\fi}

\def\feii{Fe\,{\sc ii}}

\def\cii{C\,{\sc ii}}
\def\ciii{\ifmmode {\rm C}\,{\sc iii} \else C\,{\sc iii}\fi}
\def\civ{\ifmmode {\rm C}\,{\sc iv} \else C\,{\sc iv}\fi}
\def\cv{\ifmmode {\rm C}\,{\sc v} \else C\,{\sc v}\fi}
\def\cvi{\ifmmode {\rm C}\,{\sc vi} \else C\,{\sc vi}\fi}

\def\nv{N\,{\sc v}}

\def\o5007{[O\,{\sc iii}]\,$\lambda5007$}

\def\ovi{O\,{\sc vi}}

\def\siiv{Si\,{\sc iv}}

\def\siII{Si\,{\sc ii}}

\def\siv{S\,{\sc iv}}

\def\svi{S\,{\sc vi}}

\def\feii{Fe\,{\sc ii}}

\def\aliii{Al\,{\sc iii}}

\def\o{\o}

\def\gtorder{\mathrel{\raise.3ex\hbox{$>$}\mkern-14mu
             \lower0.6ex\hbox{$\sim$}}}
\def\ltorder{\mathrel{\raise.3ex\hbox{$<$}\mkern-14mu
             \lower0.6ex\hbox{$\sim$}}}
\def\proptwid{\mathrel{\raise.3ex\hbox{$\propto$}\mkern-14mu
             \lower0.6ex\hbox{$\sim$}}}

\newcommand{\Ly}{Ly$\alpha$}

\begin{document}

\shortauthors{Dunn, et al.}
\shorttitle{\siv\ BALs}

\title{BAL Outflow Contribution to AGN Feedback: Frequency of \siv\  Outflows in the SDSS}


\author{Jay P. Dunn\altaffilmark{1},
Nahum Arav\altaffilmark{2},
Kentaro Aoki\altaffilmark{3},
Ashlee Wilkins\altaffilmark{4},
Courtney Laughlin\altaffilmark{2},
Doug Edmonds\altaffilmark{2},
\& Manuel Bautista\altaffilmark{5}}

\altaffiltext{1}{Department of Chemistry and Physics, Augusta State University, Augusta, GA 30904: jdunn10@aug.edu}
\altaffiltext{2}{Department of Physics, Virginia Tech, Blacksburg, VA 24061: arav@vt.edu,  edmonds@vt.edu}
\altaffiltext{3}{Subaru Telescope, National Astronomical Observatory of Japan, 650 North A'ohoku Place, Hilo, HI 96720, U.S.A.: kentaro.aoki@hawaiiantel.net}
\altaffiltext{4}{Cornell University: anw37@cornell.edu}
\altaffiltext{5}{Department of Physics, Western Michigan University, Kalamazoo, MI 49008-5252, USA: manuel.bautista@wmich.edu}

\begin{abstract}

We present a study of Broad Absorption Line (BAL) quasar outflows that
show \siv~$\lambda$1063 and \siv*~$\lambda$1073 troughs. The
fractional abundance of \siv\ and \civ\ peak at similar value of the
ionization parameter, implying that they arise from the same
physical component of the outflow. 
Detection of the \siv* troughs will allow us to determine the distance 
to this gas with higher resolution and higher signal-to-noise spectra, 
therefore providing the distance and energetics of the ubiquitous \civ\ 
BAL outflows. In our bright sample of 156 SDSS quasars 14\% show \civ\
and 1.9\% \siv\ troughs, which is consistent with a fainter magnitude
sample with twice as many objects. One object in the fainter sample 
shows evidence of a broad \siv\ trough without any significant trough 
present from the excited state line, which implies that this outflow 
could be at a distance of several kpc. Given the fractions of \civ\ 
and \siv, we establish firm limits on the global covering factor on 
\siv\ that ranges from 2.8\% to 21\% (allowing for the k-correction). 
Comparison of the expected optical depth for these ions with their 
detected percentage suggests that these species arise from common 
outflows with a covering factor closer to the latter.

\end{abstract}

\keywords{quasars: absorption lines, galaxies: evolution}

\section{Introduction}

Quasar outflows are detected as blueshifted absorption troughs with
respect to the AGN's rest frame spectrum.  These outflows carry mass, 
momentum and energy into the surrounding environment, and therefore
may be important in the context of AGN feedback (Arav et al 2010). To
quantify the effect of the outflows on their environment we need to
determine their average mass flow rate ($\dot{M}$) and associated
kinetic luminosity ($\dot{E}_k$). Assuming the outflow is in the form
of a partial thin, spherical shell (see Arav et al 2010 $\S$ 5.2 and
their Fig.~4) moving with a constant radial velocity $v$ at a
distance $R$ from the central source, these quantities are given by:
\begin{equation}
\dot{M} \sim 4 \pi \mu m_p \Omega R N_H v, \ \ \ \dot{E}_k =\frac{1}{2}\dot{M}v^2,
\label{mdot}
\end{equation}
where $N_H$ is the total column density of hydrogen, $m_p$ is the mass
of a proton, $\mu$=1.4 is the plasma's mean molecular weight per proton,
and $\Omega$ is the fraction of the solid angle around the quasar occupied by the
outflow. 

Over the past few years we have built a research program to determine
$\dot{M}$ and $\dot{E}_k$ in quasar outflows
\citep{2008ApJ...681..954A,2008ApJ...688..108K,2009ApJ...706..525M,2010ApJ...709..611D,2010ApJ...713...25B}. 
Targeting objects that show absorption troughs from excited and/or
meta-stable levels, we were able to obtain measurements of $N_H$ and
$R$ to an accuracy of $\sim30\%$.  The outflow's velocity is
determined simply and accurately from its blueshift in the spectrum,
which leaves $\Omega$ as the major uncertain quantity in equation
(\ref{mdot}). No information regarding $\Omega$ can be obtained from a
single object's spectrum, as we only see the absorption along the line
of sight. Therefore, our inference regarding $\Omega$ is statistical
in nature. Essentially, we use the detection rate of objects with
outflows among all quasars as an estimate for $\Omega$.  (See
discussion in Dunn et al 2010; Arav et al 2010 and \S~5 here.)

A significant uncertainty in our case stems from the fact that we can
only determine distances to outflows that show troughs from excited
and metastable levels. For most of our objects, the spectra cover such
transitions only from singly ionized species (e.g., \feii\ and \siII).
However, the majority of outflows only show troughs from high
ionization species (HiBALs) where in both optical
\citep{2003AJ....125.1784H} and IR \citep{2008ApJ....672..108D}
surveys, 20\% of all quasars are inferred to have \civ\ BALs. In
optical surveys low ionization BALs (LoBALs) are quite rare, only
0.5\% of the SDSS quasars show low ionization BALs (Trump et al 2006).
However, recent work (Dai et al 2010) have shown that in IR surveys
(that are much less sensitive to obscuration) 4\% of all quasars are
LoBALs. While these findings considerably narrow the gap between the 
two populations, we still need to address the question regarding the 
distance and energetics of both outflow manifestations. Are the LoBALs 
results representative for the HiBALs, which are majority of the outflows? 

There are good indirect arguments that outflows which
exhibit  LoBALs are regular \civ\ outflows viewed
through a high opacity line-of-sight that allows for the formation of
singly ionized troughs. One plausible example would be when the
outflows see an ionizing continuum that is filtered by the edge of
the putative AGN torus (Hall et al 2003; Dunn et al 2010, Arav et al
2010). In that case, their $\Omega$ and $R$ will be the same as that 
of the \civ\ BALs.  Alternatively, the lower detection rate can
be explained as LoBALs being a different kind of an outflow 
with $\Omega\simeq 0.04$ on average and possibly no correspondence 
with the $R$ of HiBAL outflows. 

A direct way to advance on this issue is to observe excited state
troughs from higher ionization species. Thus, a simple comparison
with the ubiquitous \civ\ is feasible and we circumvent the ambiguity
due to the preferred ionization state of the outflow. Such troughs 
appear at wavelengths short-ward of \Lya\ but have been rarely 
analyzed in the literature due to blending with the thick Lyman 
Forest absorption at high redshifts.  

Here we target the \siv/\siv*~$\lambda\lambda$1062.66,1072.97 lines
(the \siv*$\lambda1072.97$ arises from the metastable 3$s^2$3$p$
$^2$P$^o_{3/2}$ level). Absorption troughs due to \siv\ were seen
previously in BALQSOs PG~0946+301
\citep{1999ApJ...516...27A,2001ApJ....561..118A} and WPVS007 
\citep{2009ApJ...701..176L}, as well as in the Seyfert 1
galaxy NGC 4151 \citep{2006ApJS..167..161K}.  The fractional abundance
of \siv\ and \civ\ peak at similar value of the ionization parameter,
implying that they arise from the same physical component of the
outflow (see \S~3 and Fig. 3). Therefore, measuring $N_H$ and $R$ for
these outflows will will yield less model-dependent measurements for
$\dot{M}$ and $\dot{E}_k$ in the majority of outflows. Here we take 
the first step in determining the $\dot{M}$ and $\dot{E}_k$ for these
outflows by establishing limits on $\Omega$.

The plan of the paper is as follows: In \S~2, we describe our Sloan
Digital Sky Survey (SDSS) sample selection, outline the detection
methods and present the detected \siv\ outflows. In \S~3
we elaborate on the relationship between the detection rate of \siv\
vs. that of \civ, \siiv, and \cii\ troughs in quasar outflows. In \S~4,
we discuss our results and their implications with regards to $\Omega$.
Finally, we summarize our findings in \S5.

\section{Survey}

\subsection{Criteria for Detecting \siv\ Absorption Outflows in High Redshift Quasars}

Our sample takes advantage of the wealth of archived data from the
Sloan Digital Sky Survey (SDSS) conducted with a 2.5-m telescope at
Apache Point Observatory (APO) \citep{2000AJ...120.1579Y}. We search
for quasars through the SDSS data release 7 \citep[DR7;][]{2010AJ...139.2360S}
and download the calibrated spectra via the online SQL interface to
the Catalog Archive Server (CAS). We begin by introducing two key
selection criteria. Our first criterion is to include only quasars
brighter than an r' magnitude of 18. We constrain the r' magnitude
in order to keep the signal-to-noise (S/N) in the Lyman $\alpha$
forest region high enough to minimize confusion with spurious troughs.
To observe \siv $\lambda$1063,
given the 3800 \AA\ short wavelength limit of the SDSS, the second
criterion is that the redshift of the objects must be z $>$ 2.8.
These two limits result in a total of 156 quasars (after
excluding objects misidentified as quasars based on spectral
properties).

We search the data first for \civ\ absorption because: a) we do not
expect to find absorption from \siv\ unless the ubiquitous \civ\
trough is present, b) the \civ\ spectral region is free of \Ly\ forest
contamination, and c) it is the natural benchmark for assessing the
overall percentage of outflows in quasar. Our search criteria are: the
full width at half maximum (FWHM) of the absorption must be greater
than 500 km s$^{-1}$ and an outflow velocity between 0 and $-$7000 km
s$^{-1}$. The first criterion allows us to select against
associated and intervening \civ\ systems, and at the same time, yields a
wide enough \siv\ trough that can be differentiated from \Lya\ forest
troughs (see below). The second criterion prevents confusion due to
overlap of the
\siv\ troughs with absorption troughs arising from the \ovi\ doublet
($\lambda\lambda$1032, 1038).

At a redshift of z$\sim$3 we expect the number of \Ly\ absorption
systems in any given sight-line to a quasar, within our wavelength
range of (1000 to 1216) \AA, to be $\sim$150
\citep{1994ApJS...91...1B}. However, we can distinguish intrinsic
\siv\ absorption systems of the quasar from the \Ly\ forest absorbers
with the following criteria: 1) The coincidence in velocity with \civ,
and even more so with absorption troughs from \siiv\ 
(see section 2.2 and Fig~1), this is our primary method of discerning 
the origin of the absorbing gas. 2) The velocity width of the absorption 
trough. This criterion relies on the actual trough widths. Typically, 
\Ly\ forest absorbers have FWHM$\lesssim$25 km s$^{-1}$
\citep{1999MNRAS.310..289O,2008ApJ...679..194D}. The spectral
resolution of SDSS is $\sim$150 km s$^{-1}$, therefore the \Lya\
forest absorbers are not resolved. However, a FWHM of 150 km s$^{-1}$
is more than a factor of 3 smaller than our FWHM cutoff. Thus, only
chance alignments of several \Ly\ absorbers with the \civ\ absorption
in velocity space can be misidentified. 3) Our final criterion is the
presence of a trough from the excited state transition \siv*
$\lambda$1073, covering a similar velocity range. Kinematic agreement
between troughs from \siv* $\lambda$1073 and \siv\ $\lambda$1063 with
troughs not contaminated by \Ly\ forest lines (e.g., \civ, \siiv)
makes false detections due to \Ly\ lines significantly less likely.

\pagebreak

\begin{deluxetable}{lrccccc}
\tablecolumns{7}
\tablewidth{0.95\textwidth}
\tabletypesize{\normalsize}
\tablecaption{ Quasars with \civ\ Outflows in Our Sample}
\tablehead{
\colhead{SDSS Name} &
\colhead{Plate-MJD-Fiber$^a$} &
\colhead{r' Mag} &
\colhead{z$^b$} &
\colhead{\siv$^c$} &
\colhead{\siiv$^c$} & 
\colhead{\cii$^c$} \\
}
\startdata
J083535.69+212240.22 & 53349-1929-041 &  17.89 &  3.116 &   &   &   \\
J084401.95+050357.92 & 52650-1188-464 &  17.88 &  3.346 &   & y & y \\
J100841.22+362319.36 & 52993-1426-041 &  17.68 &  3.125 &   &   &   \\
J102009.99+104002.79 & 52999-1597-040 &  17.77 &  3.167 &   &   &   \\
J102025.28+334633.42 & 53442-1955-522 &  17.96 &  2.940 &   & y & y \\
J103419.71+191222.28 & 53770-2376-139 &  17.26 &  3.160 &   & y &   \\
J105101.21+153226.88 & 53852-2483-257 &  17.93 &  2.820 &   & y &   \\
J112258.77+164540.31 & 54176-2499-308 &  17.78 &  3.031 &   & y & y \\
J113008.19+535419.90 & 52707-1014-211 &  17.82 &  3.059 &   & y & y \\
J114323.71+193448.06 & 54180-2509-273 &  17.74 &  3.347 &   & y & y \\
J114548.38+393746.70 & 53442-1997-450 &  17.93 &  3.105 & y & y &   \\
J115023.58+281907.50 & 53793-2223-376 &  16.96 &  3.132 &   & y &   \\
J120934.53+553745.70 & 52707-1019-350 &  17.83 &  3.559 &   &   &   \\
J131048.17+361557.75 & 53799-2016-541 &  17.88 &  3.395 &   &   &   \\
J131927.57+445656.54 & 53089-1376-380 &  17.55 &  2.978 &   & y &   \\
J131912.40+534720.58 & 52724-1041-271 &  17.85 &  3.091 &   & y &   \\
J134722.83+465428.58 & 52736-1284-140 &  17.97 &  2.933 & y & y & y \\
J141321.05+092204.91 & 53794-1810-516 &  17.75 &  3.345 &   &   &   \\
J145125.31+144136.03 & 54242-2750-415 &  17.85 &  3.102 &   & y & y \\
J150332.18+364118.06 & 52819-1352-194 &  17.85 &  3.263 &   & y & y \\
J150923.38+243243.30 & 53820-2155-383 &  17.49 &  3.072 & y & y & y \\
J151352.52+085555.78 & 53876-1719-173 &  17.50 &  2.904 &   & y &   \\
J152553.89+513649.28 & 52378-0795-245 &  16.84 &  2.882 &   &   &   \\
J164219.89+445124.01 & 52051-0629-463 &  17.88 &  2.882 &   & y & y \\
\enddata
\normalsize

\tablenotetext{a}{The corresponding SDSS plate number, modified julian date, and 
fiber number for the object.}

\tablenotetext{b}{SDSS redshift of the object.}

\tablenotetext{c}{Absorption troughs detected at a similar velocity to \civ.}

\end{deluxetable}

\clearpage

\subsection{Quasars with \siv\ Absorption Troughs}

Given the above selection criteria, we find that of the 156 quasars,
25 have \civ\ troughs with FWHM $>$ 500 km$^{-1}$ or $\sim$16\% of the
total (see Table~1). We note though that 3 of these show very weak
\nv\ troughs and strong \siII\ troughs and are likely associated or
intervening absorption systems and not outflows from the quasar. We 
omit these 3 objects from 
further discussion, which yields a more conservative value for the 
fraction of \civ\ outflows of 14\%. We also examined the spectra for 
outflow troughs from \siiv\ and \cii. In the survey, 17 of the 
objects that show 
\civ\ outflow troughs have corresponding \siiv\ outflow troughs or 
11\% from the full sample. There are 13 objects that show \cii\ 
absorption troughs. We discuss the detection percentage of all 
the ions in Section 3.

Of the 156 objects with the proper search criteria, we find 3 robust
detections of \siv\ and \siv* absorption that appear to be kinematically 
associated with the \civ\ and \siiv\ troughs. One other object, 
SDSS~J1319+4456, appears to have a trough from \siv\ with a FWHM of 
approximately 510 km s$^{-1}$, with no associated \siv* trough. 
These less robust features warranted further scrutiny, which led us 
to obtain a higher resolution spectrum using the Subaru telescope. We 
find that this case is a superposition of 3 \Lya\ forest troughs and we
omit it from the \siv\ sample. The other 3 objects are supported by
significantly broader \siv\ troughs (too broad for the superposition of 
2 or 3 \Lya\ troughs) kinematically matching \siv* troughs, which 
makes the possibility of false detections highly unlikely.

For trough measurement 
techniques and values for the outflows in the remaining three objects, 
see Appendix A. We present the far-UV and near-UV regions of the SDSS 
spectra for the three \siv\ detections in Figure \ref{f1}. To illustrate 
the existence of \siv\ and \siv* troughs, we 
include in Figure \ref{f1} a scaled apparent optical depth template
($I = e^{-\tau_{app}}$, where $I$ is the normalized intensity) of the
\siiv\ $\lambda$1403 trough. We do so for both the \siv~$\lambda$1063
and \siv* $\lambda$1073 lines, where we allow no freedom in the 
wavelength position as it corresponds exactly to the velocity structure 
of the \siiv\ trough. We use a single scaling factor for each \siiv\
$\tau_{app}$ template to match the data. These three objects give a
conservative fraction of 1.9\% for \siv\ outflows in this sample.

We use the \siiv\ 1403 trough as a template for three reasons: a) the 
larger separation of the \siiv\ doublet (2000 km s$^{-1}$) compared to 
\civ\ (500 km s$^{-1}$) produces well separated \siiv\ troughs in most 
objects, which is not the case for the \civ\ doublet; b) as we will 
show in Section 3, \civ\ is expected to be highly saturated when \siv\ 
is detected, while \siiv\ is less saturated; c) \siiv\ 1403 is the weaker 
of the \siiv\ doublet lines (by a factor of 2) and therefore is the 
least saturated.

All three \siv\ objects show absorption troughs from \siiv, \nv, \ovi,
\Lya\ and \Lyb.  Two of the three objects show absorption from
\aliii\ and two from \cii\ (SDSS~J1145+3937 shows
neither). SDSS~J1509+2432 also shows other low ionization species
(e.g, \feii\ $\lambda$1608 and \siII\ $\lambda$1260) and an excited
\siII~$\lambda$1265 trough.  
One particular object in our sample,
SDSS~J1347+4654 has potentially saturated \siv\
absorption lines since it has similar residual
intensity in both the resonance and excited state troughs.  
This can only occur if the lines are either saturated or if the 
electron number density of the outflowing gas is finely tuned to 
10$^{4.7}$ cm$^{-3}$ (the value where the relative level populations 
of \siv\ and \siv* are
equal for T=$10^4$ K), which makes the former scenario more
likely. Relative level populations for the two \siv\ energy levels
were computed from equation 3.27 in \cite{2006agna.book...O}, which
considers only collisional and radiative transitions. Atomic data for
these computations were taken from version 2 of the XSTAR Atomic
Database, described by \cite{2001ApJS..134..139B}.

\noindent
\begin{rotate}
\begin{figure}
 \noindent
 \hspace{-1.4in}
 \begin{minipage}[t]{4.0cm}
  \vspace{-2.5cm}
  \includegraphics[width=11.6cm,height=15.0cm]{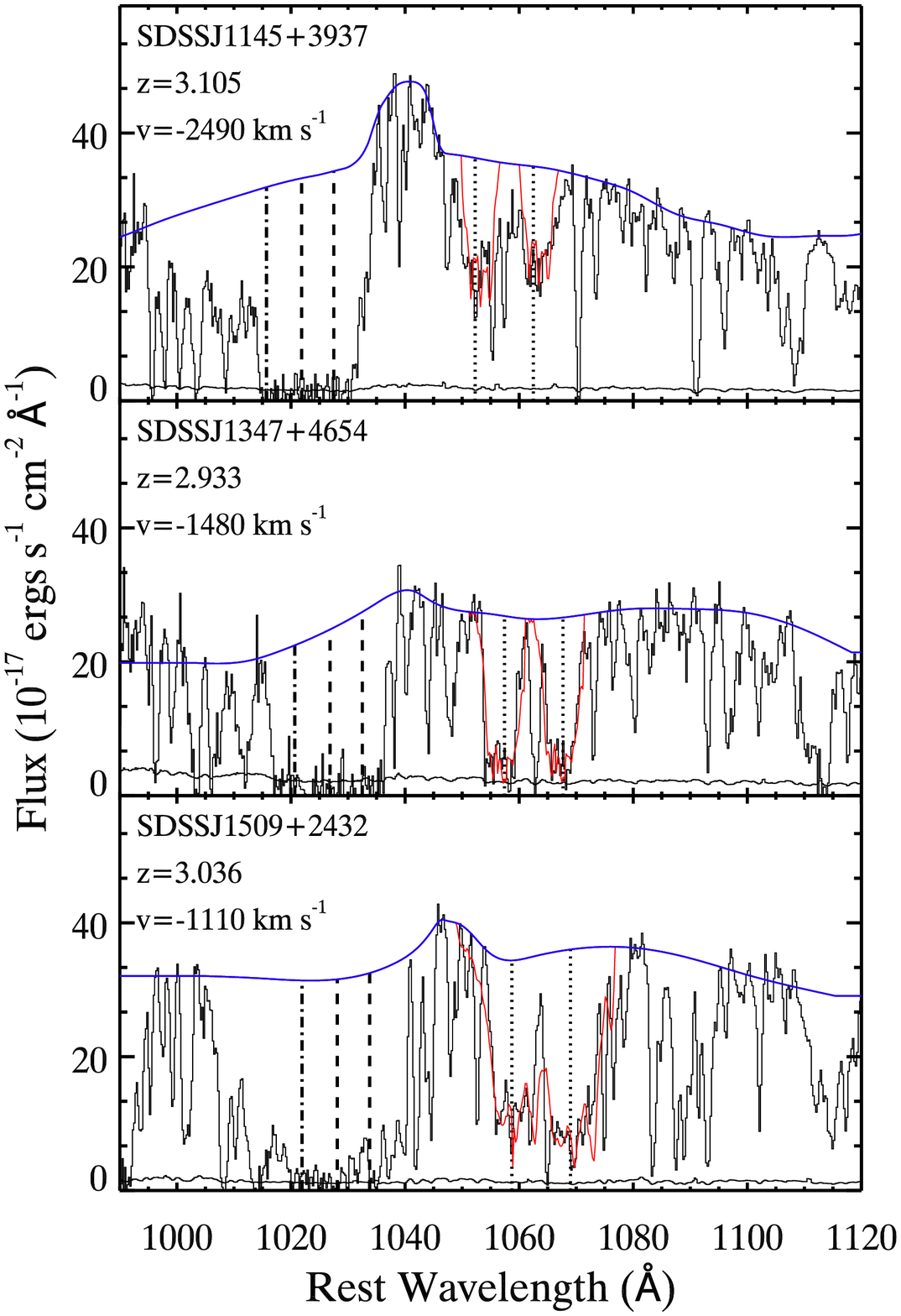}
 \end{minipage}
 \hspace{2.3in} 
 \begin{minipage}[t]{4.0cm}
  \vspace{-2.5cm}
  \includegraphics[width=11.6cm,height=15.0cm]{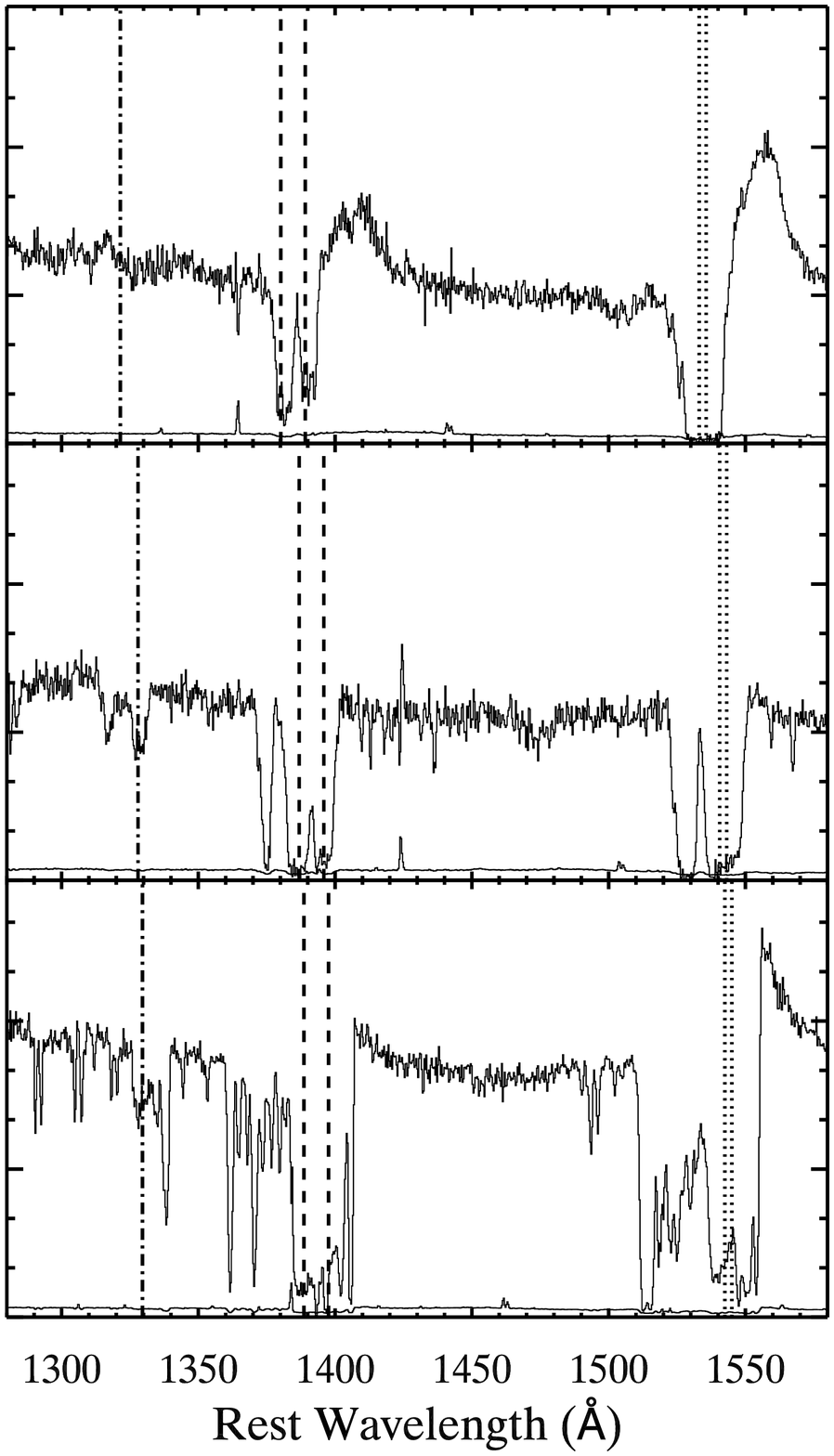}
 \end{minipage}
\caption{\scriptsize {\bf Left:} Far-UV spectra for the 3 objects that
show \siv\ and \siv* absorption. The object name, redshift and central
trough velocity are listed at the upper left of each panel.  The
dotted vertical lines mark the location of the \siv\ and \siv*
absorption troughs, while dashed lines mark the position of the
respective \ovi\ doublet $\lambda\lambda$1032,1038 positions and the
dashed-dotted lines the position of Lyman $\beta$ $\lambda$1026. The
solid blue curve shows a spline fit approximation of the continuum in
the region. In red we show a scaled optical depth template of the
\siiv\ $\lambda$1403 trough, placed at the expected position of
the \siv~$\lambda$1063 and \siv*~$\lambda$1073 lines (see text).  {\bf
Right:} Near-UV spectra for the same objects. The \civ\ troughs from
$\lambda\lambda$1548,1551 are marked with dotted lines, the \siiv\
doublet $\lambda\lambda$1394,1403 with a dashed lines, and the
expected position of the \cii\ $\lambda$1335 line, with a
dashed-dotted line.}
\label{f1}
\end{figure}
\end{rotate}

\subsection{Expanding the Search to Fainter Objects}

One concern with the above sample is basing conclusions on only three 
detections of \siv. In order to evaluate whether this result is 
representative, we also examine a fainter range in r' magnitude, 
from 18 to 18.5. This 
magnitude range provides a sample of 348 objects with the proper 
redshift and velocity criteria. Of these 348 objects, we find 8 
firm detections and 5 
weaker detections of \siv\ and \siv*. Thus, for this fainter sample, 
the detection percentage is between 2.3\% and 3.7\%. This detection 
rate is consistent with the one we find in the brighter sample.

\begin{figure}[!h]
  \centering \includegraphics[angle=0,width=0.5\textwidth]
  {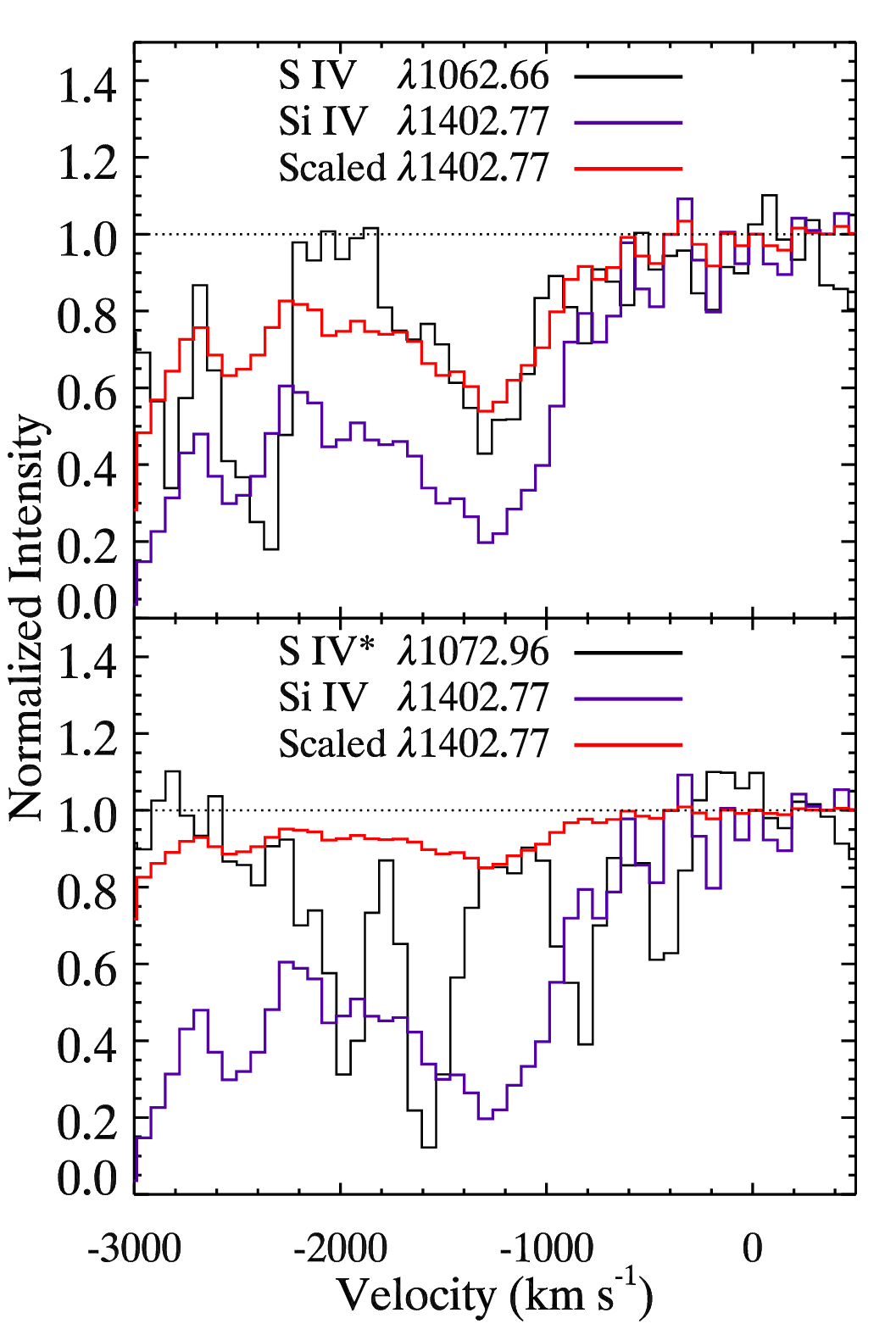}\\
  \caption[]
  {\footnotesize SDSSJ0746+3014. Top: trough profile
match between \siiv\ and \siv. Bottom: 
same as top but for the \siv* line, showing  no,  or very weak, \siv* trough using the  \siiv\
template.}
  \label{f4}
\end{figure}

In addition, in the fainter sample of objects, we find one object
(SDSSJ0746+3014) that shows evidence for a trough from the \siv\
$\lambda$1063 without the presence of the \siv* $\lambda$1073 line
(see Figure \ref{f4}).  This object shows no signs of troughs from low
ionization species (e.g., \cii, \siII). We note that the \civ\ in this
object is significantly broader but is consistent with the velocity
range of the \siv\ and \siiv\ absorption. Due to a non-detection of the
excited state line, it is plausible that that this outflow is situated 
at a distance several kpc from the central source.

\section{Comparison with detection percentage of other ionic species}

\subsection{\civ}

Quasar absorption outflows are usually identified by the appearance of
absorption from the \civ\ doublet ($\lambda\lambda$1548,1551) in their
spectrum \citep{1991ApJ...373...23W,1999ApJ...516..750C}. The ease of
\civ\ detection is due to: the abundance of carbon, the strong
oscillator strengths of these lines, the relatively low redshift in
which the \civ\ doublet can be detected from ground based
observations, and the relatively high ionization parameter ($U_H$, see
Section 5) common in UV outflows.  The expected detection ratio
between \civ\ and \siv\ is directly related to the optical depth
ratio of their lines, which depends on the relative abundances
($\chi$) of the elements , the ionic fraction of each species as a
function of $U_H$ [hereafter: $F_{\mathrm{ion}}(U_H)$], and the ratio of the
their oscillator strength ($f$) times wavelength~($\lambda$):
\begin{equation}
\frac{\tau_{\mathrm{SIV}}}{\tau_{\mathrm{CIV}}}(U_H)=
\frac{\chi(\mathrm{S})}{\chi(\mathrm{C})}
\frac{F_{\mathrm{CIV}}(U_H)}{F_{\mathrm{SIV}}(U_H)}
\frac{f_{\mathrm{SIV}}}{f_{\mathrm{CIV}}}
\frac{\lambda_{\mathrm{SIV}}}{\lambda_{\mathrm{CIV}}}=
\frac{1}{16}
\frac{F_{\mathrm{CIV}}(U_H)}{F_{\mathrm{SIV}}(U_H)}
\frac{0.05}{0.19}
\frac{1063}{1548}=
0.011\frac{F_{\mathrm{CIV}}(U_H)}{F_{\mathrm{SIV}}(U_H)}
\label{eq:tau}
\end{equation}
This is a limit on the detection ratio due to issues connected with
the line-of-sight covering of the background source (see section~3.2).

\begin{figure}[!h]
  \centering \includegraphics[angle=90,width=0.5\textwidth]
  {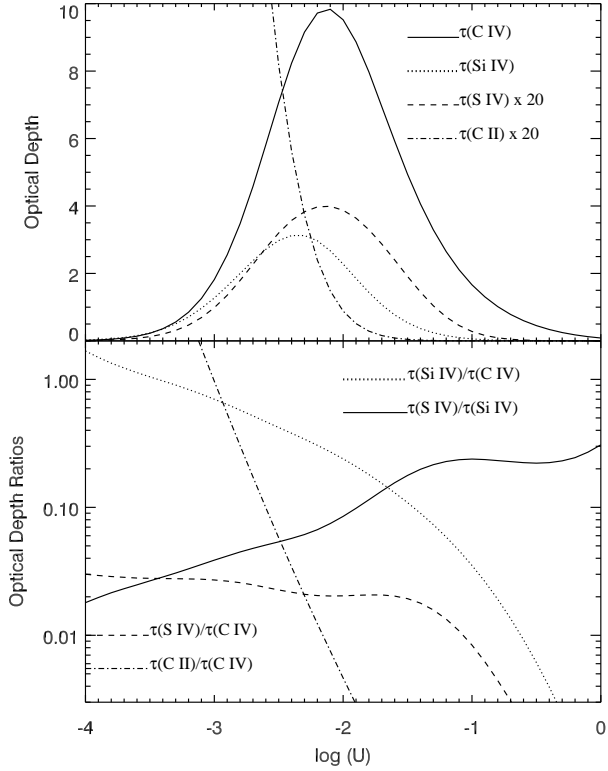}\\
  \caption[] {\footnotesize Expected optical depth for \civ, \siiv,
  \siv, and \cii\ lines, from equation (\ref{eq:tau}), assuming solar
  abundances (Lodders et al. 2003) and normalizing to
  $\tau_{max}$(civ)=10.  We used an oscillator strength of 0.05 for
  \siv~$\lambda$1063 \citep{2002MNRAS.333..885H}, 0.19 for 
  \civ~$\lambda$1548, 0.52 for \siv~$\lambda$1393 and 0.13 
  for \cii~$\lambda$1334 \citep{1995all..book.....K}. Ionization 
  fractions were calculated  with the photoionization code CLOUDY 
  using a \citet{1987ApJ...323..456M} SED.}
  \label{f3}
\end{figure}

Figure \ref{f3} shows the relative optical depth for \civ\ and \siv\
(as well as for \siiv\ and \cii) as a function of $U_H$. The shapes of
the optical depth curves as a function of $U_H$ are singularly
dependent on $F_{\mathrm{ion}}(U_H)$ (all the other variables in
equation \ref{eq:tau} simply scale the ratio by a single factor). We
observe that the relative ionization fractions for \siv\ and \civ\ are
roughly constant, peaking near log~$U_H$=$-$2.1. This behavior is
relatively insensitive to changes in the spectral energy distribution
(SED). To demonstrate this, we also calculate the ionization fractions
with a ``UV-Soft'' SED (see Dunn et al. 2010) and find only small
changes (less than 0.1 dex) for this significantly softer SED (see
online Figure 3b).  Therefore, the absorption troughs from \siv\ and
\civ\ are likely to arise from the same physical region of any given
outflow. We note that the models shown in Figure 3 are optically thin 
to all bound-free transitions, where we used a small total column 
density ($N_H$ = $10^{17}$ cm$^{-2}$) to guarantee this result in 
all possible cases.

\begin{figure}[!h]
  \centering \includegraphics[angle=90,width=0.5\textwidth]
  {fig3a.ps}\\
  \caption[] {\footnotesize This is the same as Figure 3 but
  using a ``UV-Soft'' SED (Dunn et al. 2010). This figure demonstrates
  that the ionization fraction ratio of \civ\ to \siv\ is relatively
  insensitive to changes in the SED. {\bf This figure will only appear in 
  the online version of the paper.}}
  \label{f3a}
\end{figure}

Our calculated ratio of expected optical depth depends linearly on the
S/C abundance ratio, which here we adopt as solar ($\approx$1/16; Lodders 
et al. 2003).  Carbon is mostly produced in low and intermediate mass AGB
stars, while sulfur is an $\alpha$ element and is produced in supernova 
explosions from massive stars ($M > 8 M_\odot$). Observations of Galactic 
halo stars confirm that the sulfur
abundance follows that of silicon and magnesium \citep[e.g.,][and
references therein]{2007A&A...469..319N}.  Thus, models of galactic
chemical evolution vs. metalicity
\citep[e.g.,][]{2006ApJ...653.1145K} predict variations of less than a
factor of 3 for S/C, with respect to the solar abundance ratio.

The shallowest \siv\ trough in our sample has a residual intensity of
$I_r$ = 0.36 (see Figure 1 and Table 2), and therefore an apparent
optical depth of $\tau_{app}$$\approx$1. If we assume that we
detect all \siv\ troughs in quasars with $\tau_{app}$(\siv) $>$ 1.0
(deep enough to be detected in the \Ly\ forest) and that the covering
fraction ratio of the two ions is similar (see Section 3.2), 
then from Figure \ref{f3} it is evident that $\tau$(\civ) $\gtorder 50$, 
which is their optical depth ratio. In Section
3.1, we showed that in our sample the frequency of \civ\ detection is
14\% and that of \siv\ is 1.9\%, which yields a ratio of \siv\ to
\civ\ detection of 1/7 or approximately 14\%. The large disparity in 
the optical depth
ratio explains why we do not detect \siv\ absorption troughs in every
object with a \civ\ trough and simply implies that $\tau$(\civ)
$\ltorder 50$ in $\sim$86\% of the \civ\ objects.

\subsection{Absorption Material Distribution}

One additional factor that we cannot easily quantify that would affect
the number of expected detections is the average line-of-sight (LOS) 
covering. In cases where there is LOS partial covering of the background 
source, the troughs of \siv\ will be shallower for the same optical depth. 
Thus, we would expect to detect fewer \siv\ absorption systems compared 
to \civ\ systems when the LOS covering factor ($C_f$) is small. 
Equation 2 describes the ratio of optical depth between \siv\ and \civ,
assuming complete covering. This requires a correction due to the ratio of 
the average LOS covering ($<C_f>$) between \siv\ and \civ\ if they are
different, which can be the case between ions (Arav et al. 1996; Hamann 
et al. 2001), even those that form under similar conditions. There are 
two approaches to determine $<C_f>$ for a sample of objects.

The first method to determine $<C_f>$ is to examine separated doublets
(Hamann et al. 1997) and calculate the covering factors for each object. 
However, the \civ\ doublets are blended due to our selection criteria 
and the \siv\ line is not a member of a doublet, which precludes this 
method for both ionic species. The second approach is to measure the 
average residual intensity for known saturated lines. In other words, 
the LOS covering is defined by $<C_f>$=1--$<I_{min}>$ 
\citep{2003ARA&A..41..117C}, where $I_{min}$ is the minimum 
normalized intensity in a saturated trough. The depths of the \civ\ 
absorption troughs for our sample yield $<C_f>$~$\approx$~0.80. 
We cannot guarantee an accurate determination of  $<C_f>$ for the \siv\ 
lines because only one object shows troughs with signs of saturation; 
$C_f$~$\approx$~0.82 for that case. If we assume the other two objects 
are also saturated, then $<C_f>$ is 0.71 for \siv. If the troughs
in these two objects are not saturated then this becomes a lower limit.
In either case, the difference between the $<C_f>$ for \siv\ and \civ\
is not large ($\sim$10--20\%).

The caveat with the value of $<C_f>$=0.71 for \siv\ is that this 
value can be significantly smaller. For the case where $<C_f>$ is 
different for the two ions, the main qualitative difference is that 
there are more \siv\ systems our survey will not detect. Therefore, 
our detection fraction becomes a stringent lower limit on the number 
of systems with \siv\ (we discuss the implications of this in Section 
4.1). This is because $C_f$ for \civ\ is always larger that that of 
\siv\ (as evident both in this sample and in general as \siv\ is a 
much rarer species compared to \civ; e.g., Arav et al 1999 [pg0946]). 
For any case where $C_f$ of \siv\ is smaller than $\sim$0.5, we lose 
the ability to discern \siv\ troughs from the \Lya\ forest. Thus, 
when $<C_f>$ is significantly smaller than 1.0, we will miss more such 
outflows in our SDSS survey than in cases where $C_f(C IV)\simeq 
C_f(S IV)\sim1$ and our detection fraction is a lower limit. Such 
systems may reveal themselves in data with higher spectral resolution 
and S/N.

\subsection{Comparison to Outflows with \siiv\ and \cii, and Synthesis}

In Section 3.1, we found that 11\% of the objects in our sample have
detectable \siiv\ troughs. Comparing the detection rate of \siv\ with
\siiv, the fraction of objects with \siiv\ for which we detect \siv\
is approximately 17\%. Figure \ref{f3} shows that $F_{\mathrm{SiIV}}(U_H)$ 
peaks at lower $U_H$ values than that of \civ\ and \siv\  as a function of 
$U_H$. Therefore, the predicted ratio of 
$\tau_{\mathrm{SiIV}}/\tau_{\mathrm{SIV}}$ changes from 25 to 5 as $\log(U_H)$
varies from --3 to --1.3, and has a value of 13 for $\log(U_H)=-2.1$, where
$F_{\mathrm{SIV}}(U_H)$ is at the maximum. Similarly, the
predicted $\tau_{\mathrm{SiIV}}/\tau_{\mathrm{CIV}}$ changes from 0.7 to 0.07 as
$\log(U_H)$ varies from --3 to --1.3 (we limit $-3<\log(U_H)<-1.3$,
since beyond this range $\tau_{\mathrm{SIV}}$ falls below 25\% of its maximum
value).

Both expected behaviors are consistent with the observed detection
rates in our sample. The predicted $\tau_{\mathrm{SiIV}}$ is smaller than that
of \civ, therefore fewer objects should show \siiv\ troughs compared
to \civ\ ones.  Whereas, the predicted $\tau_{\mathrm{SiIV}}$ is larger than
that of \siv, therefore more objects should show \siiv\ troughs
compared to \siv\ ones. This explanation also holds
semi-quantitatively; for the $U_H$ range of interest,
$\tau_{\mathrm{SiIV}}/\tau_{\mathrm{SIV}}$ is significantly larger than
$\tau_{\mathrm{CIV}}/\tau_{\mathrm{SiIV}}$ (by a factor ranging from 18 to 1, for
$-3<\log(U_H)<-1.3$). For this $U_H$ range we therefore expect the
fraction of \siiv\ detections to be in between, but closer to that of
\civ\ than to \siv;  in good agreement with our results of sample
detection percentages: 14\% for \civ, 11\% for \siiv, and 1.9\% for
\siv.

\cii\ outflow troughs are detected in 13 of our sample objects (8.3\%). 
Figure \ref{f3} shows that $F_{\mathrm{CII}}(U_H)$ peaks at much lower 
values that that of \siiv\ and \civ\ and drops quickly over the 
$-3<\log(U_H)<-1.3$. Combining information for the detection of all four 
ions with the expected optical depth of their troughs and assuming that
the plasma is optically thin to bound-free transitions (see Figure 3), 
a simple toy model 
can explain the difference in detection percentage: \\
1) All four ionic troughs come from the same outflows\\
2) All outflows have $\log(U_H)\sim -2.4$ \\
3) There is a wide distribution of $\tau_{\mathrm{CIV}}$ 
with $\sim$14\% of the outflows having $\tau_{\mathrm{CIV}}>50$.\\
Assumption 2) is probably the weakest, as a range of $U_H$
values is expected. Nonetheless, the implication of this toy model is
that with the same $\Omega$ for all four ions we can explain the
differing detection percentage using only the expected optical
depth of the lines.

\section{Discussion}

\subsection{Global Covering for Quasar Absorption Outflows}

As noted in the Introduction, the mass flux and kinetic luminosity of
an outflow is linearly dependent on the solid angle subtended by the
wind as viewed from the quasar ($\equiv 4\pi \Omega$). Because there is 
no direct spectroscopic determination for $\Omega$, the common method is
to statistically determine an average global covering ($C_G$) via the
detection fraction of absorbers in spectroscopic surveys and use it as
an approximation of $\Omega$ (see \S~3.2 for discussion of partial 
line-of-sight covering). For example, as mentioned in the Introduction, 
Hewitt \& Foltz (2003) and Dai et al (2008) find that 
\civ\ broad absorption lines (BALs) appear in approximately 20\% of all 
quasars. This can be interpreted as either $\Omega~\approx~C_G\sim$0.2 
in every quasar, or that $\sim$ 20\% of all quasars have their sky 
completely covered ($\Omega$~=~1), or that all quasars have $C_G\sim$1 
for 20\% of their duty cycle.  For the outflows' integrated influence 
over time for an ensemble of quasars, all three interpretations yield 
similar amounts of injected mass and energy.

Traditionally, the global covering factor for a specific type of 
outflow (e.g., outflows with troughs from \civ) is determined 
statistically via the product of the average LOS covering factor 
($<C_f>$) and the fraction of absorbers detected ($f$; Crenshaw 
et al. 2003). The values for $<C_f>$ between \siv\ and \civ\ are
plausibly similar as we show in Section 3.2 (given the assumptions
outlined). Furthermore, if we are missing detections of \siv\ due
to low $C_f$ in objects (as explained in Section 3.2), 
then the actual detection fraction $f$ is larger. Thus, regardless 
of the uncertainty in $<C_f>$, the global covering for \siv\ does
not drastically affect the $C_G$ (see Section 4.2).

Much research has been devoted to finding $f$ and thus $C_G$
for various ions in outflows. 
\citet{1999ApJ...516..750C} examined \civ\ absorption in a sample of 
low-z AGN observed with the {\it Hubble Space Telescope (HST)} and 
found that outflows in the UV in these objects were common, appearing 
in approximately 50\% of the objects in their sample.
\citet{2007AJ....134.1061D} found a similar detection rate for low-z 
AGN using absorption lines from \ovi\ observed with the {\it Far 
Ultraviolet Spectroscopic Explorer}. In higher redshift objects, 
\citet{2007ApJ...665..990G} searching through spectra from the 
Sloan Digital Sky Survey (SDSS) confirmed the 
\citet{2003AJ....125.1784H} result. 

\subsection{Global Covering for \siv\ Outflows}

In section 2.2, we showed that the conservative detection rate of \siv\
outflows is 1.9\%. 
Because we are observing SIV troughs in the optical, we need to apply 
a k-correction, which compensates for differences in the SED in the far 
UV. The factor for the k-correction is 22/15 as found by Hewett \& Folz 
(2003), which also applies to our current survey because the SDSS 
spectra cover the same rest wavelength range for these objects. Using 
the new value, we can directly compare the optical detection rate to 
the actual fraction of quasars with CIV BALs. The resulting 2.8\%
can be used as a stringent lower limit on $C_G$ of \siv\ outflows. The
upper limit is again that the global covering is equal to \civ, with
$C_G\approx$0.20. As we showed in Section 3, the detection percentage
of \civ, \siiv, \cii\ and \siv\ can all be explained straightforwardly
by the expected optical depth for the lines of each species assuming
that they arise from the same outflow and have similar $C_{f}$, which
is plausible given the small range for the ratio between \siv\ and 
\civ. Therefore, it is probable that \civ\ and \siv\ troughs arise 
from the same outflows and thus have the same $C_G$ and $\Omega$.

It is also plausible that the $N_H$ for both outflows' manifestations
is similar and the cases where we do not detect \siv\ troughs simply have 
a higher ionization parameter. Since the fraction of both \civ\ and
\siv\ drops for $\log(U_H)>-2.1$, the large difference in optical
depth can explain why the \siv\ troughs disappear while we would still
detect \civ\ (see Fig.~3). Indeed, many quasar outflows are known to have 
a higher $U_H$ value, e.g., $\log(U_H)\sim-0.8$ for the outflow seen in 
quasar J212329.46--005052.9 (Hamann et al 2010) and $\log(U_H\sim-1.4$
in the quasar FBQS~J1151+3822 \citep{2011ApJ...728...94L}. If however, 
$\log(U_H)$ for all \civ\ outflows is in a narrow range around $-2.1$,
then outflows that show \siv\ troughs will have a larger $N_H$, which will
be needed to explain the \siv\ detection.  In such a scenario, the
\siv\ outflows will tend to have higher $\dot{M}$ and $\dot{E_k}$ than
the majority of \civ\ outflows (see equation 1). 

\section{Summary and Conclusions}

We establish the detection percentage of \siv\ outflows in quasar
spectra, for troughs arising from both resonance ($\lambda$1063) and
the excited state transition ($\lambda$1073) lines.  In our bright
(brighter than r' magnitude of 18) SDSS sample of 156
quasars with $z>2.8$ (the redshift
requirement to detect \siv$\lambda$1063 in the SDSS spectra), 22 have
\civ\ outflows and 3 of these show \siv\ troughs. This yields
detection rates of 21\% for \civ\ and 2.8\% for \siv, after applying
a k-correction. We also examine
a similar sample of 348 quasars with an r' magnitude between 18 and
18.5 and find that, given the weaker signal-to-noise
ratio, the percentage of \siv\ detection is consistent
with that of the bright sample.

A detailed comparison of the expected optical depth ($\tau$) between
\civ~$\lambda$1548, \siiv~$\lambda$1393, \cii~$\lambda$1334 and
\siv~$\lambda$1063, demonstrates that the detected percentage of each
ion in our sample is in good agreement with the ratio of expected 
$\tau$ of the troughs. An analysis of $<C_f>$ for the objects with 
\siv\ troughs shows that the difference in detection fraction is
plausibly small and that a disparate $<C_f>$ between the two ions
does not strongly affect the determination of global covering. We
therefore conclude that all of these lines arise from the same 
outflows and that the fraction of solid angle subtended by them 
($\Omega$) is the same for \civ\ and \siv\ outflows, where the 
difference in detection percentage can be fully explained by the 
large expected $\tau_{CIV}/\tau_{SIV}\sim50$.

We intend to follow up all three of the objects in our bright sample
with high resolution and high signal-to-noise spectra in an effort 
to determine radial distances of the outflows. However, in Section 3, 
we find one object with a trough likely to be \siv\ whose
profile is similar to the \siiv\ trough, yet there is no evidence of a 
\siv* trough. Plausibly, this outflow could be situated at a distance of
several kpc, much like the distances we find for \feii\ and \siII\ 
outflows (see Table 3 in Arav et al 2010). Should this distance be 
confirmed with a higher quality spectrum, it 
would lend support to the assertion that the outflows where we detect
\feii\ and \siII\ troughs are the same outflows as those without but
viewed through a high opacity line-of-sight that allows for the 
formation of singly ionized species (see Hall et al 2003, Dunn et al.
2010, Arav et al, 2010). 

We acknowledge support from NSF grant AST 0837880 and from NASA LTSA
grant NAG5-12867. We would also like to thank Kirk Korista, Mike
Crenshaw and Pat Hall for their insightful suggestions and
discussions.

\appendix
\addcontentsline{tar}{section}{appendices}

\section{\civ\ and \siv\ Trough Measurements}

We measure the following trough properties from the SDSS spectra for
the \siv\ outflows: residual intensity (using our estimate of the
continuum flux level), central velocity, and the full width at half 
maximimum (FWHM) of the absorption trough, and present these in Table 
3. Measurements of the \siv\ and \siv* troughs are complicated by 
contamination from \Ly\ forest lines. We therefore outline our methods 
for determining these properties.


The first parameter we determine from the data is the central velocity 
of the outflow. We recognize that Lyman $\alpha$ intervening absorbers 
can potentially affect the centroid of the \siv\ and \svi* troughs, 
which makes these troughs less than ideal for determining the central 
velocity of the absorption. We also cannot use the \civ\ troughs due 
to our velocity width restrictions (\civ\ with FWHM larger than 500 
km s$^{-1}$), which leads to the individual troughs self-blending and 
rendering determinations of the central velocity impossible. Thus, we 
use templates of the kinematically coincident \siiv\ trough lines 
(shown in Figure 1) to determine the velocity. 
In two of the three
objects, the troughs are separated enough in velocity to allow such a 
determination. The \siiv\ troughs in SDSSJ1509+2432 suffer from 
self-blending as the FWHM of the troughs is over 2000 km s$^{-1}$.
Therefore, we resort to using the average velocity centroid of \siv\ 
and \siv* in that particular object. We do note that because we use 
the average of the two troughs the effective random shifts from \Ly\ 
forest lines should statistically be reduced.

The second parameter we measure is the residual intensity ($I_R$), 
which is defined as the normalized flux level at the deepest part 
of the trough. We employ two methods for determining the $I_R$ of 
the \siv\ troughs. First, for the broader \siv\ and \siv* troughs, 
we approximate the residual intensity ($I_R$) as the average depth 
across the bottom of the trough. This is necessary because the 
sharp variations in the trough are likely due to Lyman $\alpha$ forest 
contamination that creates a lower I$_R$ than the real \siv\ troughs. 
We estimate the error in this measurement by adding in quadrature 
the standard deviation of the points in the bottom of the trough 
and the respective average intensity error for those points. For 
the narrower \siv\ troughs (i.e., 
SDSSJ1145+3937), as well as their respective 
\civ\ troughs, we can only take the I$_R$ as the deepest point in 
the trough. To determine the error in the I$_R$ measurement we 
simply use the SDSS provided uncertainty. Finally, we use $I_R$ 
to calculate the flux at half maximum and measure the width of 
the trough. We use the uncertainty in $I_R$ to determine the 
range of acceptable half maximum within the trough. This range 
translates to an upper and lower limit for the FWHM of the trough 
determined by the walls of the trough at the extrema.

\clearpage
\begin{deluxetable}{l|lcc|lccc|lcc}
\tablecolumns{11}
\setlength{\tabcolsep}{0.04in} 
\tabletypesize{\footnotesize}
\tablecaption{Measurements of \siv\ Troughs}
\tablehead{
\multicolumn{1}{c}{} &
\multicolumn{3}{|c|}{\civ\ $\lambda\lambda$1548,1550} & 
\multicolumn{4}{|c|}{\siv\ $\lambda$1063} & 
\multicolumn{3}{|c}{\siv* $\lambda$1073} \\

\multicolumn{1}{c}{Object} &
\multicolumn{1}{|c}{FWHM} &
\multicolumn{1}{c}{I$_{r}$$^a$} &
\multicolumn{1}{c}{$\sigma_{I_r}$$^a$} &
\multicolumn{1}{|c}{FWHM} &
\multicolumn{1}{c}{Velocity} &
\multicolumn{1}{c}{I$_{r}$$^a$} &
\multicolumn{1}{c|}{$\sigma_{I_r}$$^a$} &
\multicolumn{1}{|c}{FWHM} &
\multicolumn{1}{c}{I$_{r}$$^a$} &
\multicolumn{1}{c}{$\sigma_{I_r}$$^a$} \\

\multicolumn{1}{c}{ } &
\multicolumn{1}{|c}{(km s$^{-1}$)} &
\multicolumn{1}{c}{ } &
\multicolumn{1}{c|}{ } &
\multicolumn{1}{|c}{(km s$^{-1}$)} &
\multicolumn{1}{c}{(km s$^{-1}$)} &
\multicolumn{2}{c|}{ } &
\multicolumn{1}{|c}{(km s$^{-1}$)} &
\multicolumn{2}{c}{ } \\
}
\startdata
SDSSJ1145+3937 & 4720$\pm$30 & $<$0.01& 0.03 & 1340$^{+130}_{-30}$  & $-$2490 & 0.33 & 0.05 & 1510$^{+260}_{-110}$ & 0.42 & 0.05 \\
SDSSJ1347+4654 & 3770$\pm$30 & 0.02   & 0.04 & 1660$^{+260}_{-50}$  & $-$1480 & 0.18 & 0.09 & 1610$^{+160}_{-140}$ & 0.17 & 0.08 \\
SDSSJ1509+2432 & 5240$\pm$20 & 0.05   & 0.02 & 2160$^{+260}_{-130}$ & $-$1110 & 0.36 & 0.09 & 2590$^{+80}_{-120}$  & 0.23 & 0.08 \\
\enddata
\normalsize

\tablenotetext{a}{Normalized Intensity}
\end{deluxetable}


\bibliographystyle{apj}

\bibliography{ms}

\begin{thebibliography}{29}
\expandafter\ifx\csname natexlab\endcsname\relax\def\natexlab#1{#1}\fi

\bibitem[{{Arav} {et~al.}(2001){Arav}, {de Kool}, {Korista}, {Crenshaw}, {van
  Breugel}, {Brotherton}, {Green}, {Pettini}, {Wills}, {de Vries}, {Becker},
  {Brandt}, {Green}, {Junkkarinen}, {Koratkar}, {Laor}, {Laurent-Muehleisen},
  {Mathur}, \& {Murray}}]{2001ApJ....561..118A}
{Arav}, N. {et~al.} 2001, \apj, 561, 118

\bibitem[{{Arav} {et~al.}(1999){Arav}, {Korista}, {de Kool}, {Junkkarinen}, \&
  {Begelman}}]{1999ApJ...516...27A}
{Arav}, N., {Korista}, K.~T., {de Kool}, M., {Junkkarinen}, V.~T., \&
  {Begelman}, M.~C. 1999, \apj, 516, 27

\bibitem[{{Arav} {et~al.}(2008){Arav}, {Moe}, {Costantini}, {Korista}, {Benn},
  \& {Ellison}}]{2008ApJ...681..954A}
{Arav}, N., {Moe}, M., {Costantini}, E., {Korista}, K.~T., {Benn}, C., \&
  {Ellison}, S. 2008, \apj, 681, 954

\bibitem[{{Bautista} {et~al.}(2010){Bautista}, {Dunn}, {Arav}, {Korista},
  {Moe}, \& {Benn}}]{2010ApJ...713...25B}
{Bautista}, M.~A., {Dunn}, J.~P., {Arav}, N., {Korista}, K.~T., {Moe}, M., \&
  {Benn}, C. 2010, \apj, 713, 25

\bibitem[{{Bautista} \& {Kallman}(2001)}]{2001ApJS..134..139B}
{Bautista}, M.~A., \& {Kallman}, T.~R. 2001, \apjs, 134, 139

\bibitem[{{Bechtold}(1994)}]{1994ApJS...91...1B}
{Bechtold}, J. 1994, \apjs, 91, 1

\bibitem[{{Crenshaw} {et~al.}(1999){Crenshaw}, {Kraemer}, {Boggess}, {Maran},
  {Mushotzky}, \& {Wu}}]{1999ApJ...516..750C}
{Crenshaw}, D.~M., {Kraemer}, S.~B., {Boggess}, A., {Maran}, S.~P.,
  {Mushotzky}, R.~F., \& {Wu}, C.-C. 1999, \apj, 516, 750

\bibitem[{{Crenshaw} {et~al.}(2003){Crenshaw}, {Kraemer}, \&
  {George}}]{2003ARA&A..41..117C}
{Crenshaw}, D.~M., {Kraemer}, S.~B., \& {George}, I.~M. 2003, \araa, 41, 117

\bibitem[{{Dai} {et~al.}(2008){Dai}, {Shankar}, \&
  {Sivakoff}}]{2008ApJ....672..108D}
{Dai}, X., {Shankar}, F., \& {Sivakoff}, G.~R. 2008, \apj, 672, 108

\bibitem[{{Danforth} \& {Shull}(2008)}]{2008ApJ...679..194D}
{Danforth}, C.~W., \& {Shull}, J.~M. 2008, \apj, 679, 194

\bibitem[{{Dunn} {et~al.}(2010){Dunn}, {Bautista}, {Arav}, {Moe}, {Korista},
  {Costantini}, {Benn}, {Ellison}, \& {Edmonds}}]{2010ApJ...709..611D}
{Dunn}, J.~P. {et~al.} 2010, \apj, 709, 611

\bibitem[{{Dunn} {et~al.}(2007){Dunn}, {Crenshaw}, {Kraemer}, \&
  {Gabel}}]{2007AJ....134.1061D}
{Dunn}, J.~P., {Crenshaw}, D.~M., {Kraemer}, S.~B., \& {Gabel}, J.~R. 2007,
  \aj, 134, 1061

\bibitem[{{Ganguly} {et~al.}(2007){Ganguly}, {Brotherton}, {Cales}, {Scoggins},
  {Shang}, \& {Vestergaard}}]{2007ApJ...665..990G}
{Ganguly}, R., {Brotherton}, M.~S., {Cales}, S., {Scoggins}, B., {Shang}, Z.,
  \& {Vestergaard}, M. 2007, \apj, 665, 990

\bibitem[{{Hewett} \& {Foltz}(2003)}]{2003AJ....125.1784H}
{Hewett}, P.~C., \& {Foltz}, C.~B. 2003, \aj, 125, 1784

\bibitem[{{Hibbert} {et~al.}(2002){Hibbert}, {Brage}, \&
  {Fleming}}]{2002MNRAS.333..885H}
{Hibbert}, A., {Brage}, T., \& {Fleming}, J. 2002, \mnras, 333, 885

\bibitem[{{Kobayashi} {et~al.}(2006){Kobayashi}, {Umeda}, {Nomoto}, {Tominaga},
  \& {Ohkubo}}]{2006ApJ...653.1145K}
{Kobayashi}, C., {Umeda}, H., {Nomoto}, K., {Tominaga}, N., \& {Ohkubo}, T.
  2006, \apj, 653, 1145

\bibitem[{{Korista} {et~al.}(2008){Korista}, {Bautista}, {Arav}, {Moe},
  {Costantini}, \& {Benn}}]{2008ApJ...688..108K}
{Korista}, K.~T., {Bautista}, M.~A., {Arav}, N., {Moe}, M., {Costantini}, E.,
  \& {Benn}, C. 2008, \apj, 688, 108

\bibitem[{{Kraemer} {et~al.}(2006){Kraemer}, {Crenshaw}, {Gabel}, {Kriss},
  {Netzer}, {Peterson}, {George}, {Gull}, {Hutchings}, {Mushotzky}, \&
  {Turner}}]{2006ApJS..167..161K}
{Kraemer}, S.~B. {et~al.} 2006, \apjs, 167, 161

\bibitem[{{Kurucz} \& {Bell}(1995)}]{1995all..book.....K}
{Kurucz}, R.~L., \& {Bell}, B. 1995, {Atomic line list} (Kurucz CD-ROM,
  Cambridge, MA: Smithsonian Astrophysical Observatory, |c1995, April 15, 1995)

\bibitem[{{Leighly} {et~al.}(2011){Leighly}, {Dietrich}, \&
  {Barber}}]{2011ApJ...728...94L}
{Leighly}, K.~M., {Dietrich}, M., \& {Barber}, S. 2011, \apj, 728, 94

\bibitem[{{Leighly} {et~al.}(2009){Leighly}, {Hamann}, {Casebeer}, \&
  {Grupe}}]{2009ApJ...701..176L}
{Leighly}, K.~M., {Hamann}, F., {Casebeer}, D.~A., \& {Grupe}, D. 2009, \apj,
  701, 176

\bibitem[{{Mathews} \& {Ferland}(1987)}]{1987ApJ...323..456M}
{Mathews}, W.~G., \& {Ferland}, G.~J. 1987, \apj, 323, 456

\bibitem[{{Moe} {et~al.}(2009){Moe}, {Arav}, {Bautista}, \&
  {Korista}}]{2009ApJ...706..525M}
{Moe}, M., {Arav}, N., {Bautista}, M.~A., \& {Korista}, K.~T. 2009, \apj, 706,
  525

\bibitem[{{Nissen} {et~al.}(2007){Nissen}, {Akerman}, {Asplund}, {Fabbian},
  {Kerber}, {Kaufl}, \& {Pettini}}]{2007A&A...469..319N}
{Nissen}, P.~E., {Akerman}, C., {Asplund}, M., {Fabbian}, D., {Kerber}, F.,
  {Kaufl}, H.~U., \& {Pettini}, M. 2007, \aap, 469, 319

\bibitem[{{Osterbrock} \& {Ferland}(2006)}]{2006agna.book...O}
{Osterbrock}, D.~E., \& {Ferland}, G.~J. 2006, {Astrophysics of gaseous nebulae
  and active galactic nuclei}, ed. {Osterbrock, D.~E.~\& Ferland, G.~J.}

\bibitem[{{Outram} {et~al.}(1999){Outram}, {Chaffee}, \&
  {Carswell}}]{1999MNRAS.310..289O}
{Outram}, P.~J., {Chaffee}, F.~H., \& {Carswell}, R.~F. 1999, \mnras, 310, 289

\bibitem[{{Schneider} {et~al.}(2010){Schneider}, {Richards}, {Hall}, {Strauss},
  {Anderson}, {Boroson}, {Ross}, {Shen}, {Brandt}, {Fan}, {Inada}, {Jester},
  {Knapp}, {Krawczyk}, {Thakar}, {Vanden Berk}, {Voges}, {Yanny}, {York},
  {Bahcall}, {Bizyaev}, {Blanton}, {Brewington}, {Brinkmann}, {Eisenstein},
  {Frieman}, {Fukugita}, {Gray}, {Gunn}, {Hibon}, {Ivezi{\'c}}, {Kent}, {Kron},
  {Lee}, {Lupton}, {Malanushenko}, {Malanushenko}, {Oravetz}, {Pan}, {Pier},
  {Price}, {Saxe}, {Schlegel}, {Simmons}, {Snedden}, {SubbaRao}, {Szalay}, \&
  {Weinberg}}]{2010AJ...139.2360S}
{Schneider}, D.~P. {et~al.} 2010, \aj, 139, 2360

\bibitem[{{Weymann} {et~al.}(1991){Weymann}, {Morris}, {Foltz}, \&
  {Hewett}}]{1991ApJ...373...23W}
{Weymann}, R.~J., {Morris}, S.~L., {Foltz}, C.~B., \& {Hewett}, P.~C. 1991,
  \apj, 373, 23

\bibitem[{{York} {et~al.}(2000){York}, {Adelman}, {Anderson}, {Anderson},
  {Annis}, {Bahcall}, {Bakken}, {Barkhouser}, {Bastian}, {Berman}, {Boroski},
  {Bracker}, {Briegel}, {Briggs}, {Brinkmann}, {Brunner}, {Burles}, {Carey},
  {Carr}, {Castander}, {Chen}, {Colestock}, {Connolly}, {Crocker}, {Csabai},
  {Czarapata}, {Davis}, {Doi}, {Dombeck}, {Eisenstein}, {Ellman}, {Elms},
  {Evans}, {Fan}, {Federwitz}, {Fiscelli}, {Friedman}, {Frieman}, {Fukugita},
  {Gillespie}, {Gunn}, {Gurbani}, {de Haas}, {Haldeman}, {Harris}, {Hayes},
  {Heckman}, {Hennessy}, {Hindsley}, {Holm}, {Holmgren}, {Huang}, {Hull},
  {Husby}, {Ichikawa}, {Ichikawa}, {Ivezi{\'c}}, {Kent}, {Kim}, {Kinney},
  {Klaene}, {Kleinman}, {Kleinman}, {Knapp}, {Korienek}, {Kron}, {Kunszt},
  {Lamb}, {Lee}, {Leger}, {Limmongkol}, {Lindenmeyer}, {Long}, {Loomis},
  {Loveday}, {Lucinio}, {Lupton}, {MacKinnon}, {Mannery}, {Mantsch}, {Margon},
  {McGehee}, {McKay}, {Meiksin}, {Merelli}, {Monet}, {Munn}, {Narayanan},
  {Nash}, {Neilsen}, {Neswold}, {Newberg}, {Nichol}, {Nicinski}, {Nonino},
  {Okada}, {Okamura}, {Ostriker}, {Owen}, {Pauls}, {Peoples}, {Peterson},
  {Petravick}, {Pier}, {Pope}, {Pordes}, {Prosapio}, {Rechenmacher}, {Quinn},
  {Richards}, {Richmond}, {Rivetta}, {Rockosi}, {Ruthmansdorfer}, {Sandford},
  {Schlegel}, {Schneider}, {Sekiguchi}, {Sergey}, {Shimasaku}, {Siegmund},
  {Smee}, {Smith}, {Snedden}, {Stone}, {Stoughton}, {Strauss}, {Stubbs},
  {SubbaRao}, {Szalay}, {Szapudi}, {Szokoly}, {Thakar}, {Tremonti}, {Tucker},
  {Uomoto}, {Vanden Berk}, {Vogeley}, {Waddell}, {Wang}, {Watanabe},
  {Weinberg}, {Yanny}, \& {Yasuda}}]{2000AJ...120.1579Y}
{York}, D.~G. {et~al.} 2000, \aj, 120, 1579

\end{thebibliography}

\end{document}